\documentclass[journal]{IEEEtran}
\usepackage{cite}
  \usepackage[pdftex]{graphicx}
\usepackage[caption=false,font=footnotesize]{subfig}
\usepackage{txfonts}  
\usepackage{stfloats} \fnbelowfloat
\usepackage{url}
\begin{document}
\title{Developing a Launch Package for the PEGASUS Launcher}
\author{\IEEEauthorblockN{\mbox{S.\,Hundertmark, Member, IEEE} , G.\,Vincent, D.\,Simicic}\\
\IEEEauthorblockA{French-German Research Institute, Saint Louis, France\\
}
}

\maketitle
\begin{abstract}
Railguns are capable to far exceed the muzzle energies of current naval deck guns.
Therefore one of the most promising scenario for the future application of railguns in
naval warfare is the long range artillery. Hypervelocity projectiles being propelled to
velocities above 2\,km/s reach targets at distances of 200\,km or more. At the
French-German Research Institute the PEGASUS launcher is used for investigations with
respect to this scenario. The 6\,m long barrel has a square caliber of 40\,mm. The power
supply unit is able to deliver 10\,MJ to the gun. Within this investigation, a complete
launch package is being developed and experiments are performed that aim at showing that
this package can be accelerated to velocities ranging from 2000\,m/s to 2500\,m/s. A
launch package consists out of an armature, a sabot and the projectile. The armature ensures 
the electrical contact during launch and pushes the sabot with its payload through the barrel. 
The sabot guides and protects the payload during the acceleration. At the same time the 
accelerating forces generated at the armature needs to be transferred to the projectile. 
After the launch package has left the barrel, the sabot should open and release 
its payload, the projectile into free-flight. Here the the current status
of the launch package development and results from Experiments with the PEGASUS railgun 
are presented.
\end{abstract}
\section{Introduction}
The PEGASUS railgun installation at the French-German Research Institute (ISL) is being used
for experiments in support of research for a long range artillery scenario. In future and
current modern naval ships, the electric power requirements for a large muzzle energy railgun 
can be meet \cite{mcnab,shipboard}. Compared to existing naval deck guns the large muzzle
velocities of 2\,km/s to 3\,km/s require the development of new guided hypervelocity projectiles. 
In response to this, activities for the design of such a projectile have started at ISL.
The sub-caliber projectile is embedded in a sabot when launched with a railgun.
This sabot ensures the mechanical contact to the armature, guidance through the
railgun barrel and mechanical protection. Depending on the sensitivity of the on-board
electronics of the projectile the sabot might also need to incorporate a shielding function
against electromagnetic interference from the electromagnetic fields during railgun operation. 
The assembly of the armature, the sabot and the projectile is termed launch package. After the
launch package has left the barrel, the projectile has to separate from the armature and
sabot to embark on its free flight trajectory. Work at other research labs concerning the 
development and testing of electromagnetic gun launch packages is described in the recent 
literature \cite{crawford,satapathy,kitzmiller,zielinski}. The aim of the investigation 
described in this report is to demonstrate that close to realistic projectiles can be successfully
launched with the existing railgun technology. The experience gained with the experiments
will allow to improve on the weak points of the preliminary design and to
give the groups working on the hypervelocity projectile (payload, electronics and the projectile
itself) a framework in which testing in an actual railgun environment is possible.
\section{PEGASUS Launcher Setup}
\begin{figure}[tb!]
\centering
\includegraphics[width=3.5in]{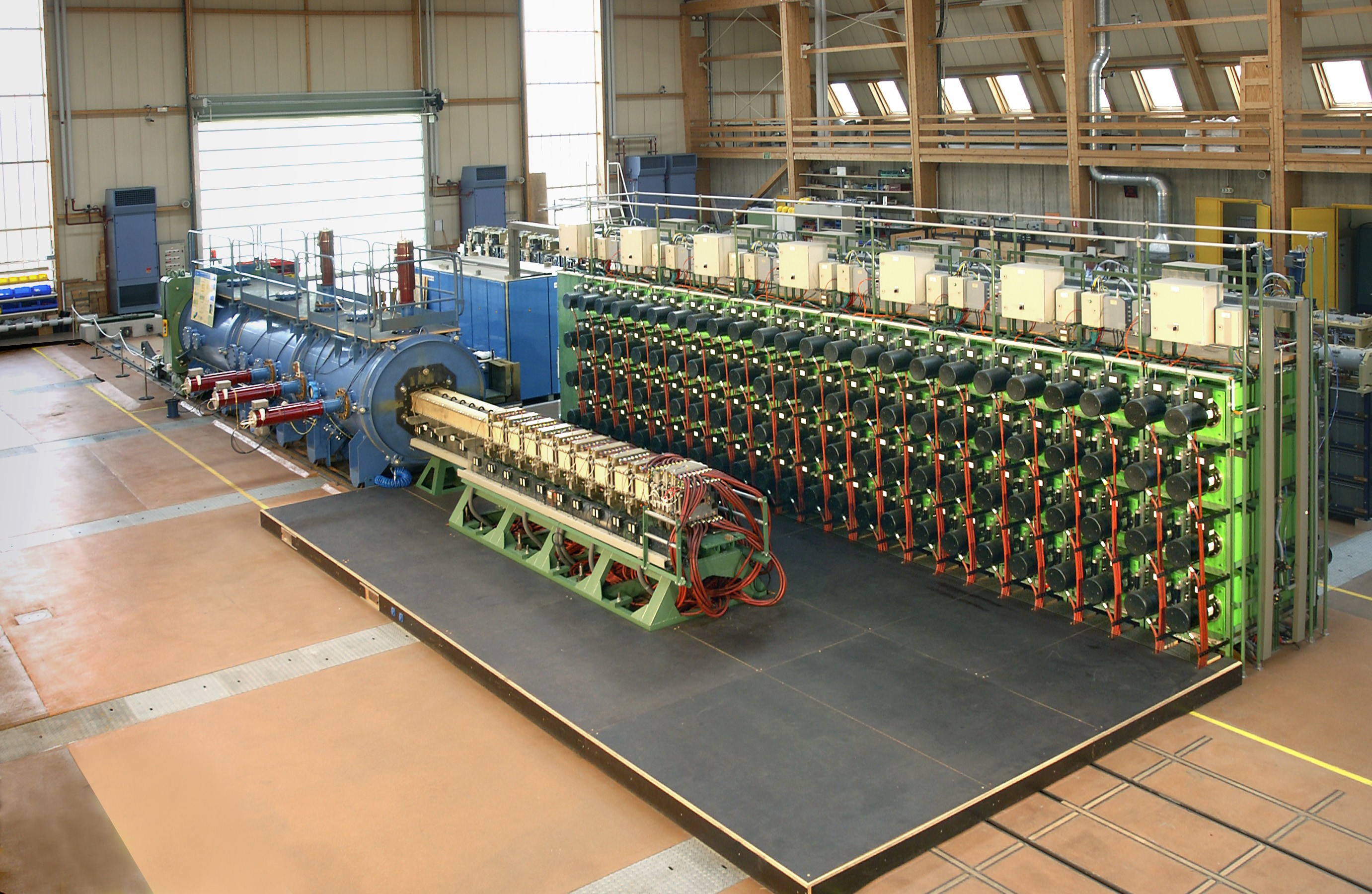}
\caption{PEGASUS railgun and its 10\,MJ power supply.}
\label{peg}
\end{figure}
The PEGASUS railgun used in these experiments is a 40\,mm square caliber and 6\,m long
railgun. It is connected to a 10\,MJ capacitor based power supply. Each of the 200 
capacitor modules can store up to 50\,kJ. It is equipped with a thyristor, a crow-bar
diode, a pulse forming coil and a trigger circuit. Each module is connected by one 
coaxial cable to current injection points distributed along the first
3.75\,m of acceleration length. This distributed energy supply (DES) scheme reduces the
ohmic losses and minimizes the residual magnetic energy in the barrel. By correlating the
release of parts of the stored energy with the progress of the armature through the barrel
a relatively flat current pulse with a large DC contribution can be generated. The
experimental setup is shown in figure \ref{peg}. To the left in the figure, the 7\,m long
catch tank is visible.  To monitor the condition of the launch package or armature/projectile, 
a high-speed camera and  several flash X-ray tubes can be mounted on the catch tank. 
Several B-dot sensors are placed along the barrel length. The signals from these sensors are used
to calculate the armature velocity. Doppler radars inside the catch tank allow for an
independent measurement of the end-velocity. In recent experiments, using c-shaped aluminum 
armatures, velocities of more than 3\,km/s were reached with an efficiency of approx. 40\%
including the power supply losses \cite{eml_1}.
\section{Hybrid Armature}
\begin{figure}[tb!]
\centering
\includegraphics[width=3.5in]{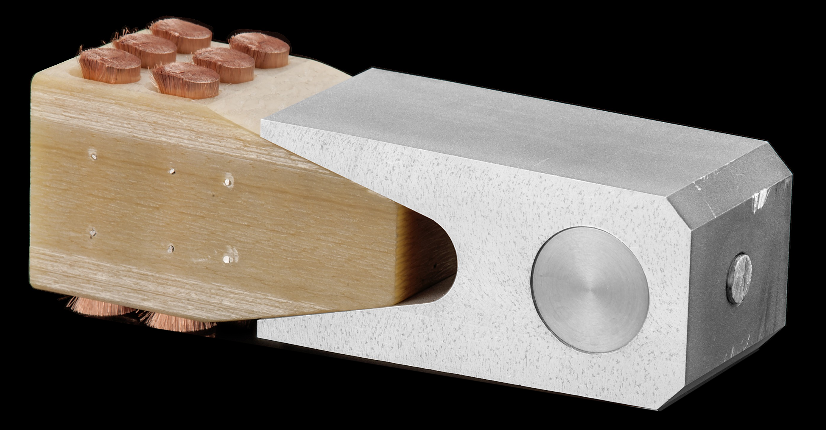}
\caption{Pushed c-shaped or hybrid armature.}
\label{hybrid}
\end{figure}
A c-shaped aluminum armature establishes electrical contact to the rail surface via its two
trailing arms. Once the current flow is established through the armature, the Lorentz 
force ensures that the arms are firmly pressed against the rail surface. During current ramp-up
after triggering the launch, this contact force can be established by over-sizing the
distance between the arm outer surfaces above the measured caliber. Unfortunately it is not 
straight forward to calculate the amount of pre-stress and the arm distance being required 
to ensure a good electrical contact behavior during the start-up phase. In addition to this, the 
PEGASUS barrel shows caliber variations after usage. It is therefore not practical to
correctly size the armature to the barrel condition. The hybrid armature as shown in
figure \ref{hybrid} is the answer to this problem. A brush equipped
glass-fiber reinforced plastic (GRP) armature (left side) is combined with a c-shaped aluminum 
armature (right side). 
The brush equipped armature is located at the rear of the combined armature and is equipped 
with a nose that fits into the spacing in between the c-shaped armature arms. There is a
small space available in between the GRP nose tip and the c-shaped throat. The working
principle of this hybrid technique is as follows: Initially the brushes of the rear
armature part carry (most of) the current and the so accelerated body pushes the
projectile forward. The inertia of the aluminum part forces the nose into the available
space and pushes the arms of the c-shaped armature apart and against the rails. At one point during
acceleration, the brushes are eroded and the electric current is forced to use the
short-circuit path through the arms of the aluminum armature. Due to the strong forces
accelerating the projectile, the GRP armature forces the legs strongly against the rails. 
After the current has moved to the c-shaped armature, this part is further accelerated and
the brush equipped part falls behind.

\subsection{Hybrid Armature Experiment}
\begin{figure}[tb!]
\centering
\includegraphics[width=3.5in]{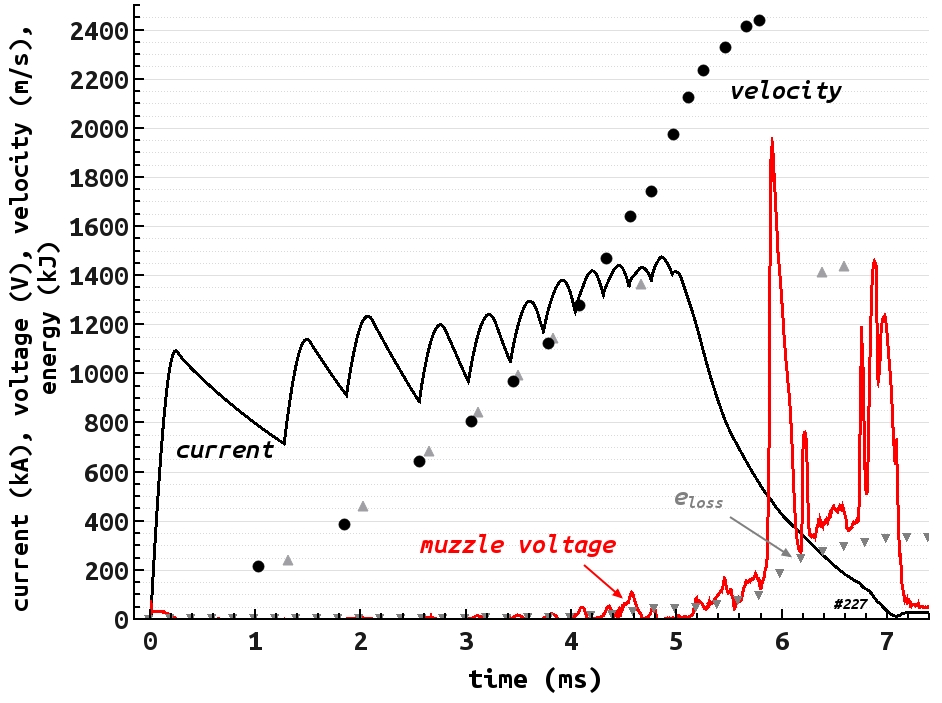}
\caption{Hybrid armature launch at 4.9\,MJ. Shown are the current, the muzzle voltage, the
velocity of the c-shaped armature (black dots), the velocity of the brush armature
(upward, gray triangles) and the energy lost at the rail/armature interface (downward,
gray triangles).}
\label{t227}
\end{figure}
Using 4.9\,MJ of electrical energy a hybrid armature projectile with a total mass of 715\,g 
was launched. The relevant launch parameters from this launch are shown in figure
\ref{t227}. Within 0.2\,ms the current rises to 1.1\,MA and than further increases to
1.5\,MA during the launch. The velocity of the aluminum part including a payload cylinder
with a combined mass of 450\,g reaches a velocity of 2450\,m/s. At about 4\,ms and a velocity of 
1450\,m/s, the c-shaped armature separates from the GRP brush armature. A launch
efficiency of 33\% is calculated by dividing the kinetic energy by the
initial energy stored in the power supply unit. This
number includes the energy losses in the power supply unit and the cables to the launcher
barrel. The muzzle voltage trace in figure \ref{t227} shows low values until shot-out at 5.8\,ms. 
The electrical contact of the armature/rail interface was excellent during the full
acceleration length. The contact losses amount to 110\,kJ, corresponding to approx. 2\%
of the initial energy.
\section{Launch Package}
\begin{figure}[b!]
\centering
\includegraphics[width=3.5in]{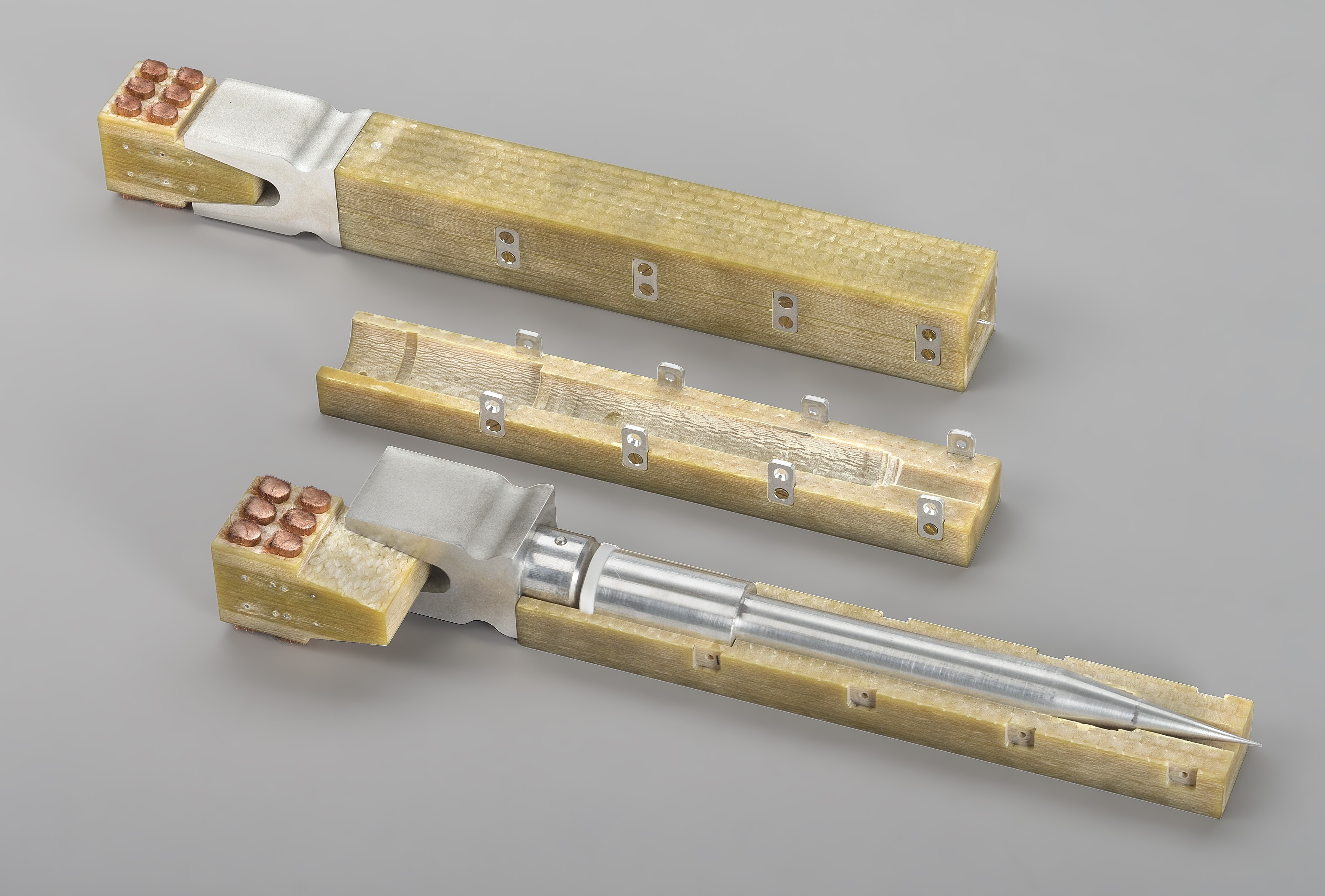}
\caption{Launch package for the PEGASUS 40\,mm barrel.}
\label{package}
\end{figure}
In the long range artillery scenario, a large caliber railgun accelerates a hypervelocity
projectile to a muzzle velocity above 2000\,m/s. To minimize the effects of the
lower, denser atmosphere on the projectile, the firing is under a steep angle. The projectile reaches
its apogee at approx. 100\,km height and descends on its ballistic trajectory until
it enters again the denser part of the atmosphere. At a height of about 10\,km, an active
correction of the flight path using information from its internal guidance and navigation system 
allows to accurately hit a target at a distance of 100\,km or more \cite{mcnab,shipboard}. 
As the flight time
from firing the gun to hitting the target is several minutes, it is conceivable that the
target data is being updated by using a radio link before the projectile reenters the atmosphere. 
The launch package developed for the 40\,mm PEGASUS barrel is shown in figure \ref{package}. It
consists out of the hybrid armature, a GRP sabot and the hypervelocity projectile model as
payload. The projectile model is simplified in the sense, that it is made fully out of
aluminum, has no wings for guidance and contains no electronics and military payload.
The c-shaped aluminum armature is equipped with a cylindrical nose that connects
the armature mechanically with the sabot. In addition to this, a pin secures the
connection between the aluminum armature and the GRP sabot. 
The GRP sabot is machined in a way that ensures a large area contact in between the
embedded  hypervelocity projectile and the sabot. In between the front of the
cylindrical nose of the armature and the rear of the projectile a white plastic damper plate 
is supposed
to dampen the mechanical shocks that come from the hard acceleration during current ramp-up 
and the not fully constant acceleration during launch. The sabot is fabricated out of two parts 
that are
weakly connected by four small aluminum pieces on each side. The nose tip of the hypervelocity
projectile protrudes out of the sabot and the central opening of the sabot front side is
designed to allow for
the opening of the sabot due to air pressure once the launch package has left the barrel.
The total mass of the launch package is approx. 1.3\,kg, with the individual masses itemized
in table \ref{tab_1}. 
Due to the large mass of the sabot, the payload-to-total-mass-ratio
is rather small (17\%).
\begin{table}
\centering\small
\begin{tabular}[tbh]{|l|r|r|}
\hline
Part & Length (mm) & Mass (g)\\
\hline
Brush armature & 67 & 235 \\
\hline
C-shaped armature &90 & 217 \\
\hline
Damper &5 & 5 \\
\hline
Sabot &260 & 590\\
\hline
Projectile &240 & 215 \\
\hline\hline
Total & 370 & 1262\\
\hline
\end{tabular}
\vspace{2mm}
\caption{Length and mass for the individual parts of the launch package.}
\label{tab_1}
\end{table}
\section{Experimental Results}
\subsection{First Launch}
\begin{figure}[b!]
\centering
\includegraphics[width=3.5in]{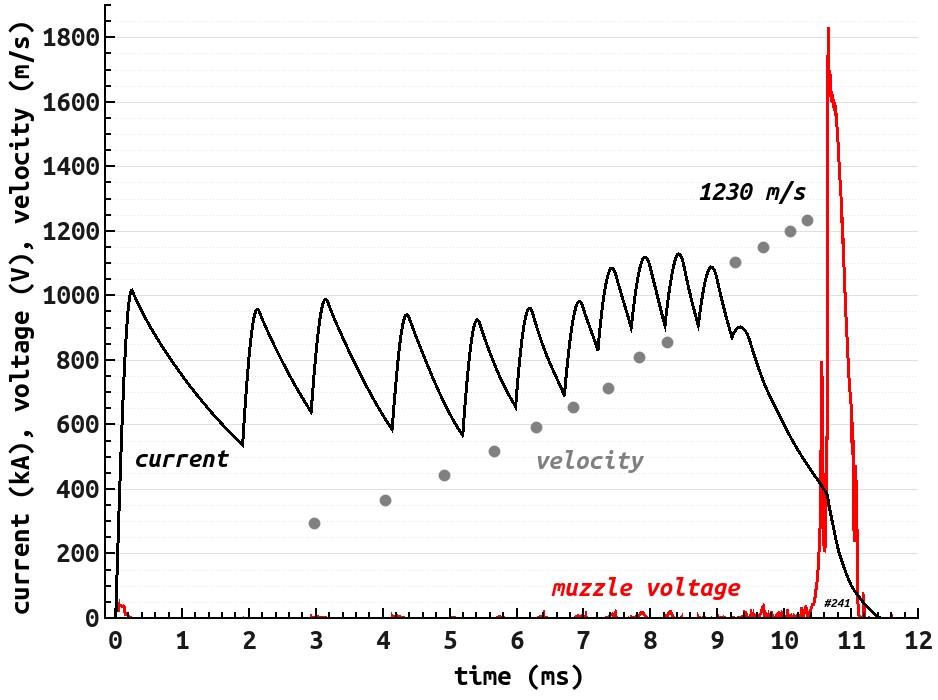}
\caption{Launch parameters for the first shot with the PEGASUS launch package.}
\label{t241}
\end{figure}
The developed launch package was used in a launch with the capacitors charged to a voltage
corresponding to 4.2\,MJ of electrical energy. Figure \ref{t241} shows the measured data 
of this shot. At an average value of 805\,kA until shot-out at 10.3\,ms, the current
reaches a peak value of 1.13\,MA. Due to the DES setup of this PEGASUS barrel, with fixed
distances in between the current injection points, shots with relatively low armature
velocity generate the strong saw-tooth pattern of the current as seen in the figure \ref{t241}.
Only at velocities which correspond to the barrel design velocity of approx. 2500\,m/s, the current
trace becomes more DC like. During the full acceleration length, the muzzle voltage is at the
lower values of the measurement range of the data-acquisition system and the energy lost at
the rail/armature interface amounts to not more than a few kilojoules. The velocity at
muzzle exit is 1230\,m/s, with an overall efficiency of 22\% for this shot. As the
separation of the GRP-brush armature from the c-shaped armature only happens very late in
the launch period (at about 8.6\,ms), both parts leave the barrel with a very short
distance in between each other. This is the cause of the double peak structure of the muzzle 
voltage trace just at 10.5\,ms.
\begin{figure}[t!]
\centering
\includegraphics[width=3.5in]{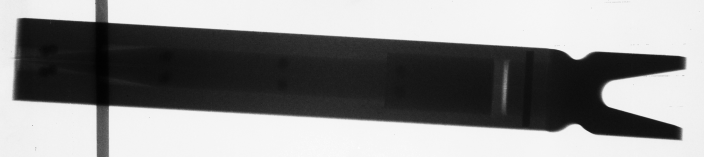}
\caption{PEGASUS launch package in free flight.}
\label{t241_xray}
\end{figure}
In figure \ref{t241_xray}, a flash x-ray picture of the launch package just after it has
left the barrel is shown. The aluminum c-shaped armature is seen in the right hand side of
the figure. At the energy used in this shot, there are no visible signs of wear of the
arms of the armature. At the time of the free flight phase when this x-ray picture was taken, 
the sabot is still closed. Visual pictures taken by a high-speed camera installed further 
down the catch tank show that the sabot opens later and frees the projectile.

\subsection{Second Launch}
\begin{figure}[b!]
\centering
\includegraphics[width=3.5in]{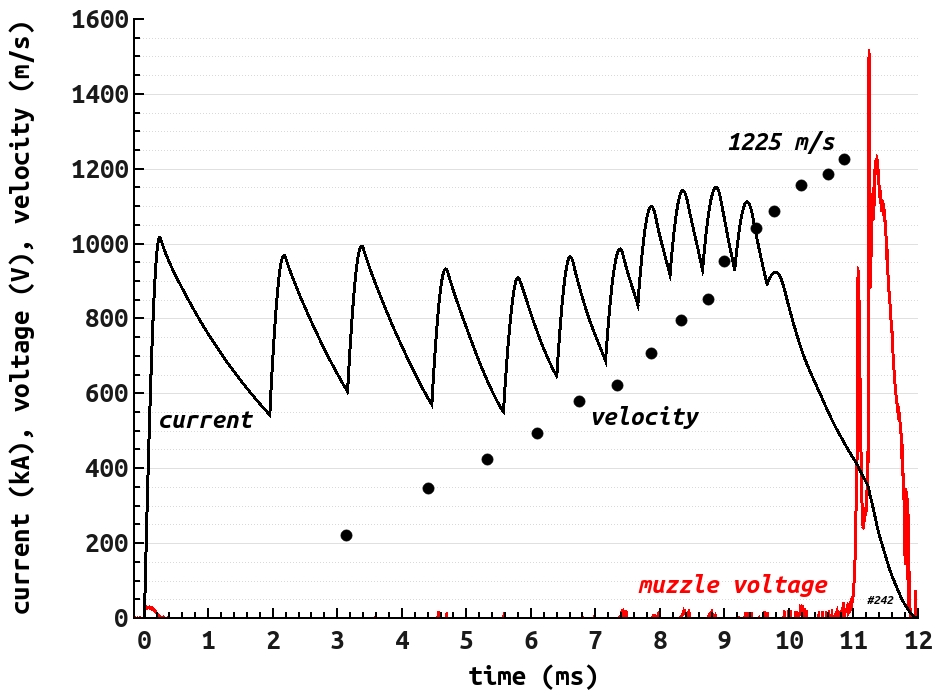}
\caption{Parameters of the second launch.}
\label{t242}
\end{figure}
\begin{figure}[t!]
\centering
\includegraphics[width=3.5in]{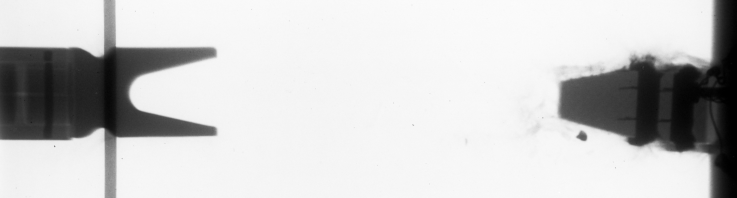}
\caption{X-ray picture of the armatures as they leave the barrel.}
\label{t242_xray_a}
\end{figure}
\begin{figure}[t!]
\centering
\includegraphics[width=3.5in]{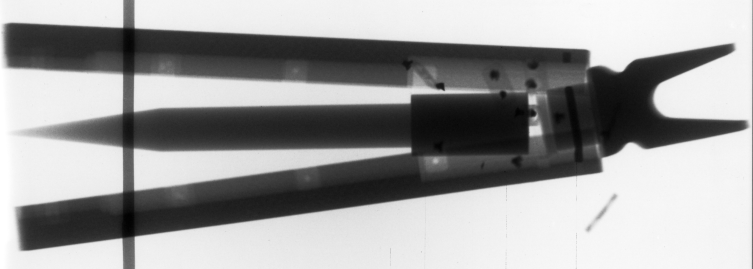}
\caption{Opening of the sabot during free flight.}
\label{t242_xray_b}
\end{figure}

With the aim to confirm the results of the first launch, a second experiment was set up. The
second launch package was prepared and launched at the same energy. The result of this
second launch is shown in figure \ref{t242}. As expected, the time development of the 
current and muzzle voltage traces are similar, as is the muzzle velocity of 1225\,m/s. The flash
x-ray picture in figure \ref{t242_xray_a} is taken just after both armatures have left the barrel.
It shows the excellent
condition of the c-shaped aluminum armature after the launch. The brushes of the brush 
equipped armature show clear signs of wear, more on the rear than on the front row. 
As in the first shot, the separation
of the rear armature from the rest happened late in the acceleration, thus the
velocity of the armatures is not very different. Clearly the current load sharing in
between the two armatures could be better balanced in future experiments, thus realizing
an earlier separation. In figure \ref{t242_xray_b} a second x-ray picture taken further down 
in the catch tank  shows the launch package during its short free flight phase. The GRP 
sabot is just in the process to open and release the hypervelocity projectile, 
validating the design of the opening mechanism.
\section{Expected Performance}
\begin{figure}[b!]
\centering
\includegraphics[width=3.5in]{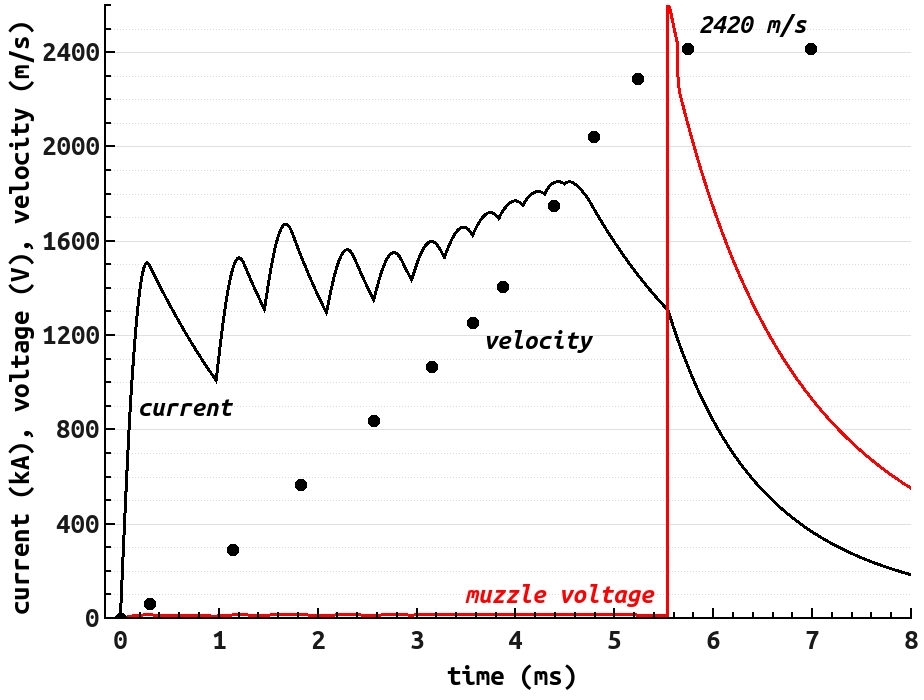}
\caption{Simulated launch behavior at 10\,MJ.}
\label{sim_10mj}
\end{figure}
For the two experiments described, the initial energy being stored in the capacitors
amounted to approx. half of the full complement. It is interesting to
investigate what velocity could be possible if the full 10\,MJ is being used. This was
investigated by a SPICE simulation. Such a simulation can give answers to
the electrical and dynamical behavior of the launcher and the projectile. Of course due to
the many uncertainties in electromagnetic launch processes at higher energies, the
simulation results are predictions that come with a certain uncertainty. This uncertainty
is hard to quantify, but from experience from comparing with already performed experiments
one can estimate the uncertainty to be of the order of 10\% to 20\%. In figure
\ref{sim_10mj} the current, muzzle voltage and velocity trace is shown. At 10\,MJ the
average current during the acceleration time amounts to 1.5\,MA, the peak value is
1.85\,MA at 4.5\,ms. Shot out is expected at 5.6\,ms. The launch package is accelerated to
a velocity of 2400\,m/s. Comparing the current trace to the two experimental traces in
figures \ref{t241} and \ref{t242} shows that the sawtooth pattern of the current 
is much less pronounced at this higher velocity. As a note of caution: it is not clear that 
this velocity can be reached with the developed launch package, as it assumes that the 
rail/armature contact does not fail during the launch process. Currently this can only be
investigated experimentally.
\section{summary}
The usage of hybrid armatures, the combination of a brush equipped and a c-shaped
aluminum armature allowed for a drastic reduction of losses at the rail/armature
interface. Building on the encouraging results with this type of armature a launch package
for the acceleration of hypervelocity projectiles was developed. The ability to accelerate
launch packages is the prerequisite for the military application of a railgun in a long
range artillery system. In experiments with the launch package developed at ISL using the PEGASUS
40\,mm railgun barrel, it was shown that the payload could be guided
through the railgun barrel and released after the package left the barrel. X-ray pictures
showed the excellent condition of the c-shaped armature and the sabot after the launch to
a velocity of 1230\,m/s. Using a simulation, the expected velocity of the projectile when
launched with an energy of 10\,MJ was evaluated to 2400\,m/s. This validates that using
the PEGASUS launcher it is possible to accelerate close to realistic models of
hypervelocity projectiles to the military relevant velocity above 2000\,m/s. 

\section*{Acknowledgment}
This research was supported by the French and German Ministries of Defense.


\begin{thebibliography}{99}
\bibitem{mcnab} 
I.\,R.~McNab,
\emph{Parameters for an Electromagnetic Naval Railgun}, 
IEEE Transactions on Magnetics, Vol.\,37, No.\,1, January 2001.

\bibitem{shipboard}
S.~Hundertmark, D.~Lancelle,
\emph{A Scenario for a Future European Shipboard Railgun},
IEEE Transactions on Plasma Science, Vol.\,43, No.\,5, May 2015.
\bibitem{crawford}
M.~Crawford, R.~Subramanian, T.~Watt, D.~Surls, D.~Motes, J.~Mallick,
D.~Barnette, S.~Satapathy, and J.~Campos,
\emph{The Design and Testing of a Large-Caliber Railgun},
IEEE Transactions on Magnetics, Vol.\,45, No.\,1, January 2009.
\bibitem{satapathy}
S.~Satapathy, I.~R.~McNab, M.~Erengil, and W.~S.~Lawhorn,
\emph{Design of an 8-MJ Integrated Launch Package},
IEEE Transactions on Magnetics, Vol.\,41, No.\,1, January 2005.
\bibitem{kitzmiller}
J.~R.~Kitzmiller, M.~D.~Driga,
\emph{An Optimized Double Ramp Integrated Launch Package Design for
Railguns},
IEEE Transactions on Magnetics, Vol.\,39, No.\,1, January 2003.
\bibitem{zielinski}
A.E.Zielinski,
\emph{Integrated Launch Package Design with Considerations for Reduced
Scale Demonstration},
ARL-TR-2315, January 2001.
\bibitem{eml_1}
S.~Hundertmark, G.~Vincent, D.~Simicic, M.~Schneider,
\emph{Increasing Launch Efficiency with the PEGASUS Launcher},
Contribution to the 18$^{th}$ EML Symposium, Wuhan, China, 24--28. October 2016. 



\end{thebibliography}
\end{document}